# Multi-Scale Convolutional Neural Network for Automated AMD Classification using Retinal OCT Images

Saman Sotoudeh-Paima[1], Ata Jodeiri[1], Fedra Hajizadeh[2,*], Hamid Soltanian-Zadeh[1,3,*]

1. Control and Intelligent Processing Center of Excellence (CIPCE), School of Electrical and Computer Engineering, College of Engineering, University of Tehran, Tehran 14399, Iran

2. Noor Ophthalmology Research Center, Noor Eye Hospital, Tehran 19686, Iran

3. Medical Image Analysis Laboratory, Departments of Radiology and Research Administration, Henry Ford Health System, Detroit, MI 48202, USA

---

* **Corresponding author:**

   Hamid Soltanian-Zadeh

   Control and Intelligent Processing Center of Excellence (CIPCE), School of Electrical and Computer Engineering, College of Engineering, University of Tehran, Tehran 14399, Iran

   Email: hszadeh@ut.ac.ir

   Tel: (+98) 912 103 9353

* **Co-corresponding author:**

   Fedra Hajizadeh

   Noor Ophthalmology Research Center, Noor Eye Hospital, Tehran 19686, Iran

   Email: fedra_hajizadeh@yahoo.com

   Tel: (+98) 912 304 5172

## Abstract

**Background and Objective:**

Age-related macular degeneration (AMD) is the most common cause of blindness in developed countries, especially in people over 60 years of age. The workload of specialists and the healthcare system in this field has increased in recent years mainly due to three reasons: 1) increased use of retinal optical coherence tomography (OCT) imaging technique, 2) prevalence of population aging worldwide, and 3) chronic nature of AMD. Recent advancements in the field of deep learning have provided a unique opportunity for the development of fully automated diagnosis frameworks. Considering the presence of AMD-related retinal pathologies in varying sizes in OCT images, our objective was to propose a multi-scale convolutional neural network (CNN) that can capture inter-scale variations and improve performance using a feature fusion strategy across convolutional blocks.

**Methods:**

Our proposed method introduces a multi-scale CNN based on the feature pyramid network (FPN) structure. This method is used for the reliable diagnosis of normal and two common clinical characteristics of dry and wet AMD, namely drusen and choroidal neovascularization (CNV). The proposed method is evaluated on the national dataset gathered at Hospital (NEH) for this study, consisting of 12649 retinal OCT images from 441 patients, and the UCSD public dataset, consisting of 108312 OCT images from 4686 patients.

**Results:**

Experimental results show the superior performance of our proposed multi-scale structure over several well-known OCT classification frameworks. This feature combination strategy has proved to be effective on all tested backbone models, with improvements ranging from 0.4% to 3.3%. In addition, gradual learning has proved to be effective in improving performance in two consecutive stages. In the first stage, the performance was boosted from 87.2% ± 2.5% to 92.0% ± 1.6% using pre-trained ImageNet weights. In the second stage, another performance boost from 92.0% ± 1.6% to 93.4% ± 1.4% was observed as a result of fine-tuning the previous model on the UCSD dataset. Lastly, generating heatmaps provided additional proof for the effectiveness of our multi-scale structure, enabling the detection of retinal pathologies appearing in different sizes.

**Conclusion:**

The promising quantitative results of the proposed architecture, along with qualitative evaluations through generating heatmaps, prove the suitability of the proposed method to be used as a screening tool in healthcare centers assisting ophthalmologists in making better diagnostic decisions.

**Keywords:**

1. **Introduction**

Age-related Macular Degeneration (AMD) is a highly prevalent retinal disorder that accounts for 8.7% of blindness globally [1]. It is the most frequent cause of blindness in developed countries, especially in people over 60, and is labeled a "priority eye disease" by the WHO [1], [2]. AMD cases fall into two general categories: dry and wet. Dry AMD accounts for 80-90% of cases. The common clinical characteristic of dry AMD is the presence of drusen, which are deposits of extracellular material that build up between the retinal pigment epithelium (RPE) and inner collagenous zone of Bruch's membrane [3], [4]. These deposits accumulate over time and lead to the damage of the RPE and subsequent loss of photoreceptor cells [5], [6]. In 10-20% of cases, patients with dry AMD develop wet AMD, in which normal blood vessels grow into the retina and leak fluid, making the retina wet. Technically, this is called CNV or choroidal neovascularization, which leads to significant visual impairment. Fig. 1 illustrates the OCT B-scans for normal, drusen, and CNV cases.

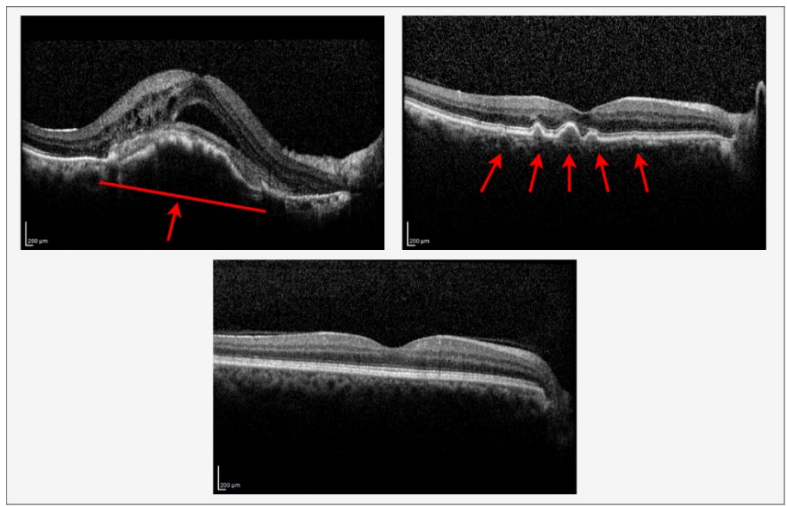

**Fig. 1.** Example OCT B-scans from the Noor Eye Hospital dataset. (a) CNV case, (b) Drusen case, (c) Normal case. Red arrows indicate the affected area in the B-scan.

The introduction of anti-angiogenesis therapy has fortunately brought about significant advancements in the management of exudative or so-called wet AMD, and intravitreal injection of anti-vascular endothelial growth factor (anti-VEGF) drugs is currently considered to be the optimal treatment for CNV [1], [7]. However, these treatments are costly and not available in all countries [1]. Moreover, any improvement is accompanied by long-term monthly intravitreal injections and uncertainty about the therapy duration and likely recurrence of CNV [7]. Thus, patient screening and early detection of AMD cases with effective diagnostic tools are critical.

Optical coherence tomography (OCT) has become the most commonly used imaging modality in ophthalmology, with more than 5 million OCTs performed in 2014 in the US Medicare population [8]. OCT is a non-invasive imaging technique that provides cross-sectional images of the macula or optic nerve head using low-coherence light [9]. Considering its non-invasiveness and ease of imaging acquisition, OCT is highly preferred by ophthalmologists for the assessment of retinal pathologies, e.g., AMD [5]. However, precise examination of multiple OCT cross-sections for each patient is a time-consuming and demanding task for ophthalmologists. Moreover, the chronic nature of AMD further increases the burden on ophthalmologists and healthcare centers. Thus, the presence of an automated computer-aided diagnosis (CAD)-based screening tool could help in prioritizing patients with respect to their condition and reducing this burden.

Therefore, in this study, we propose a novel multi-scale CNN with an FPN-based feature fusion strategy. The proposed model takes advantage of multi-scale receptive fields, enabling more accurate detection of retinal pathologies that appear in varying scales in OCT images. This method enables end-to-end training of the multi-scale model with a single CNN using a simplistic design and eliminates the need to perform preprocessing on the input data. Our experimental results on the NEH dataset published in this study and the UCSD dataset demonstrate the superior performance of our proposed methodology against several state-of-the-art retinal OCT classification frameworks. The resulting framework can also be used as a screening tool to prioritize cases depending on their condition and act as a second pair of eyes for ophthalmologists to better detect AMD-related retinal pathologies.

In the following section, we discuss the related literature on automated classification of retinal pathologies and the motivation for our proposed model.

## 2. Related Works

Numerous computerized algorithms for automated classification of retinal pathologies have been developed during recent years for preprocessing [10]–[12], classification [3], [5], [13]–[28], and segmentation [29]–[36] of OCT images. This study is focused on the classification of retinal pathologies, and in this category, studies are divided into two main branches: feature-based and deep learning-based methods. This section discusses the related literature in these two branches and reveals the motivation behind our proposed classification methodology.

### 2.1. Feature-based methods

Traditional machine learning approaches for semi/fully automatic classification of OCT images consist of three main blocks: preprocessing, feature extraction, and classifier design [37]. The preprocessing block (e.g., image denoising [11] and retinal flattening [12]) allows for the removal of unwanted or unnecessary information from the raw input data and allows the model to extract meaningful information in the following stage. Then, feature descriptors (e.g., histogram of oriented gradients (HOG) [21], [23], linear binary patterns (LBP) [27], and scale-invariant feature transform (SIFT) [12]) are employed, allowing manual extraction of features. In the end, the extracted features are fed into a classifier (e.g., random forest algorithm [22], bayesian classifier [23], and support vector machine [21], [27]) to finalize the classification process. Table 1 summarizes the previous work conducted for feature-based retinal OCT classification.

**Table 1.** Summary of previous works using feature-based methods

| Authors | Methods | Dataset | Performance Measures | Notes |
|---|---|---|---|---|
| Albarrak et al. [23] | Combine concepts of volume decomposition and LBP for feature extraction and use Bayesian classifier on the generated feature vectors | Private dataset of 140 3D OCT volumes | Accuracy: 91.4% Sensitivity: 92.4% Specificity: 90.5% | Combination of image decomposition and LBP histograms helped to form a more accurate feature descriptor for classification purposes. |
| Srinivasan et al. [21] | Proposed an algorithm that uses the multi-scale histogram of gradient descriptors as feature extractors and support vector machine as the classifier | Duke dataset [21] | Achieved an accuracy of 95.56% for patient-wise classification of normal, AMD, and DME cases | The patient is classified as normal/AMD/DME if 33% or more of the images in a volume are classified as those cases. This threshold is selected experimentally and might not be the best choice among different datasets. |
| Lemaitre et al. [27] | A classification framework with five distinctive steps was proposed | SERI private dataset [27] | The best settings achieved a sensitivity of 81.2% and specificity of 93.7% | The five steps included preprocessing (non-local means, flattening, alignment), feature detection (LBP, LBP-TOP), mapping (global, local), feature representation (histogram bag-of-words), and classification (random forest, k-NN, RBF-SVM, logistic regression, and gradient boosting). |
| Sun et al. [12] | A classification framework based on sparse coding and dictionary learning was proposed. | Duke dataset [21] + Private dataset | Achieved a patient-wise accuracy of 97.78% on the Duke dataset | A volume was appointed to a specific class (AMD, DME, or normal) by the label for the majority of the images. The average preprocessing time for a single OCT scan was 9.2 seconds which can be a limitation in real-time settings. |
| Venhuizen et al. [22] | A machine learning algorithm for automated grading of AMD severity stages was developed | European Genetic Database (EUGENDA) | The system achieved an AUROC of 0.980 with a sensitivity of 98.2% and specificity of 91.2% for high-risk AMD detection. | The algorithm showed similar performance as human observers who achieved sensitivities of 97.0% and 99.4% at specificities of 89.7% and 87.2%. |

Although machine learning approaches have proved to achieve promising results, they come with several limitations. First, manual extraction of features is a time-consuming task requiring an expert's skill, making it inefficient to collect a large and comprehensive database. Furthermore, expert interpretations might be different, leading to results that are not acceptable by other experts. This would result in models which are not generalizable to new databases.

## 2.2. Deep Learning-based methods

Deep learning (DL), a subfield of artificial intelligence (AI), has recently gained significant interest in medicine and healthcare and has been primarily applied to medical image analysis [38]. DL methods are based on representation learning, where a multi-layer neural network automatically discovers the representations needed for the classification task without any manual feature engineering, replacing the multi-block approach of traditional methods [38], [39]. Convolutional neural network (CNN) architectures have shown promising results in classifying retinal pathologies using OCT images. Table 2 summarizes the previous works conducted for automated retinal OCT classification.

**Table 2.** Summary of previous works using deep learning-based methods

| Authors | Methods | Dataset | Performance Measures | Notes |
|---|---|---|---|---|
| Lee et al. [40] | A modified version of the VGG16 CNN was used for the classification of normal and AMD cases. | Private dataset of 48312 normal and 52690 AMD macular OCT scans | Accuracy of 87.63% on the OCT level, 88.98% in the volume level, and 93.45% in the patient level | This study was the first to demonstrate the ability of deep learning models to distinguish AMD from normal OCT images. |
| Kermany et al. [20] | A transfer learning algorithm based on InceptionV3 architecture to classify CNV, DME, drusen, and normal cases | UCSD dataset [20] | Accuracy: 96.6% Sensitivity: 97.8% Specificity: 97.4% | This study demonstrated the competitive performance of the transfer learning algorithm, which eliminates the need for a highly specialized deep learning model and a dataset of millions of images. |
| Li et al. [41] | A deep transfer learning method to fine-tune the VGG16 network pre-trained on the ImageNet database | UCSD dataset [20] | Accuracy: 98.6% Sensitivity: 97.8% Specificity: 99.4% | A similar study to the one conducted by Kermany et al. [20], with the difference of using VGG16 network instead of InceptionV3 for transfer learning. |
| Kaymak et al. [42] | The original AlexNet was trained for the classification of retinal OCT pathologies. | UCSD dataset [20] | Accuracy: 97.1% Sensitivity: 99.6% Specificity: 98.4% | Due to the availability of a large OCT dataset (>100k images), AlexNet has outperformed the transfer learning method proposed by Kermany et al. [20]. |
| Serener et al. [43] | AlexNet and ResNet18 models were compared for the classification of dry and wet AMD | UCSD dataset [20] | (ResNet18- Dry AMD) Accuracy: 99.5% Sensitivity: 98.0% Specificity: 100.0% (ResNet18- Wet AMD) Accuracy: 98.8% Sensitivity: 95.6% Specificity: 99.9% | ResNet18 model outperformed the AlexNet model on both classification tasks. Further analysis demonstrated a more accurate classification of dry AMD than wet AMD. |

| Hwang et al. [7] | A deep transfer learning method for fine tuning three different architectures (VGG16, InceptionV3, ResNet50) for the classification of retinal pathologies. | Private dataset + UCSD dataset [20] | Reported accuracy on the UCSD dataset was 91.20%, 96.93%, and 95.87% for the VGG16, InceptionV3, and ResNet50 model for the classification of normal, dry AMD, and wet AMD cases | InceptionV3 model outperformed VGG16 and ResNet50 for both datasets. |
|---|---|---|---|---|
| Fang et al. [44] (JVCIR) | Iterative fusion convolutional neural network (IFCNN) method | 2nd version of the UCSD dataset [20] + MURA dataset | Reported an overall accuracy of 87.3% using five-fold cross-validation on the UCSD dataset | The proposed IFCNN method exploits the information among different convolutional layers through an iterative layer fusion strategy. |
| Huang et al. [17] | Layer guided convolutional neural network (LGCNN) | 2nd version of the UCSD dataset [20] + HUCM dataset | Reported an overall accuracy of 88.4% using five-fold cross-validation on the UCSD dataset | Retinal layer segmentation maps and two lesion-related layer information were first extracted using ReLayNet and then LGCNN was employed for integrating the extracted information for classification. |
| Rasti et al. [3] | A novel methodology based on a multi-scale convolutional mixture of expert (MCME) ensemble model | NEH dataset [3] | Reported a precision of 99.36%, recall of 99.36%, and f1-score of 99.34% on a three-class classification problem (normal, AMD, DME) | The mathematical model of the presented methodology was coupled with a new cost function based on the addition of a cross-correlation penalty term. The best accuracy is dependent on manual tuning of the loss function. Using multiple CNNs increases inference time and computational complexity. |
| Das et al. [45] | A multi-scale deep feature fusion (MDFF) approach using CNNs | 2nd version of the UCSD dataset [20] | Accuracy: 99.6% Sensitivity: 99.6% Specificity: 99.87% | Fusion of features from multiple scales can capture the inter-scale variations introducing complementary information to the classifier. In addition, no additional tuning of hyperparameters is needed (a limitation of the study conducted by Rasti et al. [3]). Using multiple CNNs increases inference time and computational complexity. |
| Thomas et al. [46] | a multi-scale CNN structure | UCSD dataset [20] | Weighted average accuracy of 99.73% for binary classification of normal vs. AMD cases | The multi-scale feature extraction architecture helps the network to create local structures of various filter sizes. |
| Fang et al. [16] (TMI) | Lesion-aware convolutional neural network (LACNN) that incorporates attention maps from a lesion detection network (LDN) | UCSD dataset [20] | Reported an overall accuracy of 90.1% using five-fold cross-validation on the UCSD dataset | Demonstrated that the detected macular lesion information can guide the network to pay more attention to discriminative features and ignore insignificant information. Usage of two separate networks (LDN+LACNN) increases computational complexity. |
| Das et al. [48] | B-scan attentive convolutional neural network (BACNN) | DUIA dataset [47] + NEH dataset [3] | Reported an overall accuracy of 90.1% on the NEH dataset and 97.1% on the DUIA dataset | The proposed methodology uses a self-attention mechanism to automatically assign appropriate weights to the clinically informative (pathological) B-scans |
| Hassan et al. [51] | Deep retinal analysis and grading framework (RAG-FW) | Duke1 [47], Duke2 [49], Duke3 [21], BIOMISA [50], and UCSD dataset [20] | Accuracy: 98.6% Sensitivity: 98.27% Specificity: 99.6% | The proposed method is a hybrid convolutional neural network (RAG-FW), employing RAG-Net that contains a segmentation and a classification unit for retinal lesion extraction and lesion-influenced grading of retinal diseases. |

This study aimed to expand the current body of work on multi-scale convolutional neural networks. Compared to the reviewed works summarized in Table 2, the main contributions of this study are: (a) feature combination among CNN blocks using FPN structure to take advantage of multi-scale receptive fields, enabling more accurate detection of pathologies appearing in different scales, (b) enabling end-to-end training with multiple scales, eliminating the need to use image pyramids and reducing computational complexity, (c) showing the robustness of the algorithm and improvement in accuracy using four famous backbone structures (VGG, ResNet, DenseNet, EfficientNet), (d) providing qualitative proof (heatmaps) supporting the usefulness of the multi-scale structure, and (e) further performance enhancement using a two-staged (gradual) learning strategy.

The rest of the sections are organized as follows: Section 3 describes the collected database and the proposed methodology, Section 4 presents the results and discussions, and Section 5 concludes this paper.

## 3. Materials and Methods

This section discusses the details of the databases used in this study and describes the proposed multi-scale CNN framework.

### 3.1. Database

The proposed method was evaluated on two separate databases. For the first database, our study used anonymized OCT images collected by the Heidelberg SD-OCT imaging system at Noor Eye Hospital, Tehran, Iran. There were no marks/features and no patient identifiers in the images.

Table 3 shows the details of the first database. All the OCT B-scans are labeled by a retinal specialist. The inclusion criteria for patient selection are having more than 50 years of age, absence of any other retinal pathology in the patient's OCT B-scans, and good image quality ($Q \geq 20$[1]). For training and comparing purposes, the worst-case condition B-scans for each volume were kept (i.e., if a patient was detected as a CNV case, only CNV-appearing B-scans were included for the training procedure), and other B-scans were eliminated from the database. Thus, among 16822 overall OCT B-scans, 12649 are used for training and testing. To enable future research on the same topic, we made the dataset available at: https://data.mendeley.com/datasets/8kt969dhx6/1

---

[1] Measured by the Heidelberg SD-OCT imaging system and provided in patients images

The second database is the University of California San Diego (UCSD) database [20]. This database consists of a train and a test set, belonging to four categories of CNV, DME, drusen, and normal. The training set contains 108312 retinal OCT images (CNV: 37206, DME: 11349, drusen: 8617, normal: 51140), and the testing set contains 1000 retinal OCT images (250 from each class). This dataset is available at: https://data.mendeley.com/datasets/rscbjbr9sj

**Table 3.** Specifications for the Noor Eye Hospital database.

| Class | # Patients | # Eyes (OD, OS) | # OCT B-Scans |
|---|---|---|---|
| Normal | 120 | 187 (95, 92) | 5667 |
| Drusen | 160 | 194 (112, 82) | 3742 |
| CNV | 161 | 173 (83, 90) | 3240 |
| Total | 441 | 554 (290, 264) | 12649 |
| Total (Before Elimination) | 441 | 554 (290, 264) | 16822 |

### 3.2. The proposed method

One major challenge with medical images is that regions of interest (ROIs) appear in varying scales. Thus, different-sized receptive fields would be needed in order to detect retinal pathologies. To achieve such an architecture, we propose a multi-scale CNN structure based on the FPN design [52]. FPN's main applications are in object detection and semantic segmentation. However, we have modified their structure so that we would be able to take advantage of their multi-scale architecture in our classification problem.

In FPN, earlier feature maps in a convolutional model have high resolution and weak semantics. On the other hand, later feature maps have low resolution and strong semantics. The goal of using an FPN-based structure is to leverage the pyramidal shape of a CNN's feature hierarchy in order to create a model with strong semantics at all scales. Using FPN-based architecture to achieve this goal, we merged high-dimensional, semantically weak feature maps through top-down pathways and lateral connections with low-dimensional, semantically strong ones. The resulting model has strong semantics at all scales and is capable of extracting features in different sizes. Extracting multi-scale feature maps and merging them using a single CNN reduces computational and memory costs.

To summarize, the benefits of having a multi-scale CNN based on FPN structure are two-fold:

1. Unlike featurized image pyramids where multiple input images with varying scales are utilized, our proposed architecture works with a single input image, reducing computational costs.

2. Unlike several previous models that used multiple CNN models running in parallel, the proposed multi-scale structure uses a single CNN to extract different-sized features and merges them to reach the overall classification result.

### 3.3. Multi-Scale CNN Structure

This Subsection provides a detailed description of the proposed structure. The multi-scale structure can be used with any off-the-shelf CNN architecture (VGG, ResNet, DenseNet, etc.) as the backbone. In this paper, we used VGG16 as the backbone network as it had the best performance when combined with the FPN structure and named it FPN-VGG16. Fig. 2 illustrates the structure of this model.

The FPN-VGG16 structure is composed of three main components: (a) encoder, (b) feature fusion using FPN architecture, and (c) classifier. The encoder part is responsible for creating the pyramidal feature hierarchy and could be selected from a wide variety of famous deep learning networks, such as VGGNets [53], ResNets [54], DenseNets [55], EfficientNets [56], etc. The feature fusion section is based on the FPN structure. In this section, the output feature maps at different scales $i$ ($i \in \{1,2,3,4,...\}$ dependant on the number of feature scales with $i = 1$ starting from earlier blocks going to $i = n$ corresponding to the last block) are first convolved with a $1 \times 1$ filter of size 256 to equalize the effect of each scale and enable addition operation with feature maps from the previous (finer) scale. Then, the resulting feature maps are merged with the ones at scale $i+1$ through addition. Let $X^i$ be the output feature map with size $(x, x, 256)$ that gone through convolution with a $1 \times 1$ filter. To enable addition operation of a coarser-resolution convolutional block of size $(x/2, x/2, 256)$ with $X^i$, $X^{i+1}_{(x/2,x/2,256)}$ need to be upsampled (named as $\hat{X}^{i+1}_{(x,x,256)}$). The last convolutional block would be transferred to the next layer without any change. This can be formulated as:

$$Y^i = \begin{cases} X^i_{(x,x,256)} + Upsampled\left(X^{i+1}_{(x/2,x/2,256)}\right) = X^i_{(x,x,256)} + \hat{X}^{i+1}_{(x,x,256)} & , i \in \{1,2,...,n-1\} \\ X^i_{(x,x,256)} & , i \in \{n\} \end{cases}$$

Where $Y^i$ is the output from the addition operation at block $i$. Two $3 \times 3$ convolutional layers are appended at the end of the feature fusion stage to extract the semantically strong scale-representative feature maps. Then, global average pooling layers were used to convert the extracted feature maps of size $(x, x, 256)$ to a feature vector of size 256 [57]. One major advantage of the global average pooling layer is that no parameters need to be optimized; thus, no overfitting will occur because of using this layer. Also, as explained in [57], the global average pooling makes the network more robust to spatial translations of the input as it sums out the spatial information. To form the final feature vector, we concatenate features from all scales, which gives us a 1280-unit feature vector. Then, this vector is connected to a dense layer with a size of 512 through a fully connected layer. To reduce the model's overfitting, we used a dropout layer with a value of 0.5. In the end, a softmax output layer gives the probability of classes for each input image. The number of output neurons was dependent on the number of classes in each dataset (three neurons for the NEH dataset with classes of normal, drusen, and CNV and four neurons for the UCSD dataset with classes of normal, drusen, AMD, and DME).

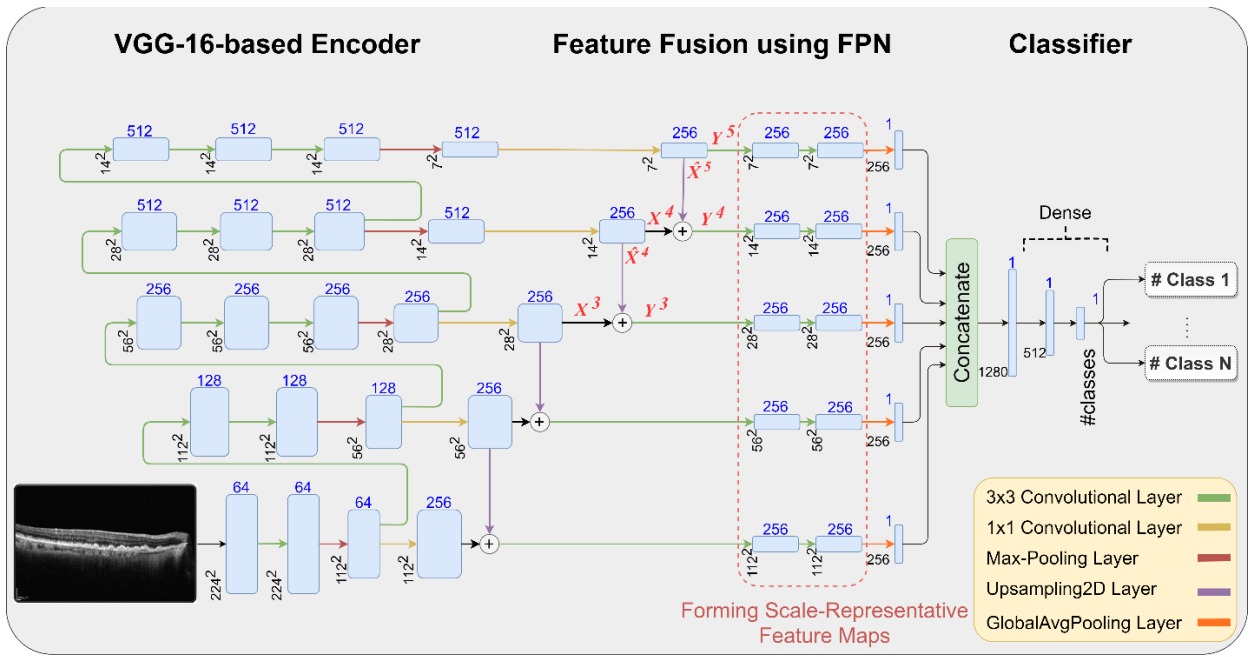

**Fig. 2.** FPN-VGG16 model structure. The model consists of three main parts: (a) encoder, (b) feature fusion using FPN architecture, and (c) classifier. The model's encoder section encodes the input image into several feature blocks. The FPN-based feature fusion section takes the input from the encoder's output and fuses them to improve the semantic representation of the model. The classifier is responsible for providing the class probabilities for each input image.

### 3.4. Experimental Setup

First, image intensities are normalized with a mean of zero and a standard deviation of one. Then, to reduce the computational load, all images are resized to $224 \times 224$. To improve variability in data and generalizability of the proposed model, we used data augmentation techniques, such as random rotation and shearing, brightness change, zoom change, and horizontal flipping. Table 4 gives the specifications of our data augmentation strategy.

**Table 4.** Specifications of the data augmentation used in this study.

| Augmentation Type | Value |
|---|---|
| Rotation range | $\pm 15\ degrees$ |
| Shear range | $\pm 5\ degrees$ |
| Brightness range | $\pm 20\%$ |
| Zoom range | $\pm 20\%$ |
| Horizontal flip | $True$ |

In this study, five-fold cross-validation at the patient level was utilized as an unbiased estimator of the models' performance in order to evaluate and compare the model's performance against baselines and previous studies. The method splits the patient data into five folds, trains the model on four of those subsets, and tests it on the subset not used for training. This process is repeated five times, with a different subset reserved for testing each time. In each run, 20% of the data within the training set is selected for validation and optimization, helping the model achieve the best generalization performance and preventing the overfitting problem. After five runs, every patient is selected exactly once as the testing set.

All the networks were trained end-to-end using an Adam optimizer and a batch size of 16. Weighted categorical cross-entropy (CCE) loss function was used to compensate for the imbalanced data distribution in both datasets. Table 5 shows the class weights for both datasets. Learning rate decay was used as a strategy to help the model achieve the best performance or least loss value during optimization. This strategy starts training the model with a large learning rate and slowly decays it until a local minimum is obtained. Early stopping was another strategy used to regularize the deep learning model. This strategy monitors the performance of the model after each epoch during training on a validation set and terminates the training process when no decline in loss value is observed. In this study, we started with a learning of 1e-4 and reduced it by half after each epoch that validation loss did not improve (decrease). Early stopping was set to ten epochs in this study, so the training process terminates if validation loss does not decrease after ten consecutive epochs.

**Table 5.** Class weights for the Noor Eye Hospital and UCSD datasets.

| Dataset | Classes | # B-Scans | Class Weights |
|---|---|---|---|
| Noor Eye Hospital Dataset | CNV | 3240 | 0.26 |
| | Drusen | 3742 | 0.29 |
| | Normal | 5667 | 0.45 |
| | Total | 12649 | 1 |
| UCSD Dataset | CNV | 37206 | 0.34 |
| | DME | 11349 | 0.11 |
| | Drusen | 8617 | 0.08 |
| | Normal | 51140 | 0.47 |
| | Total | 108312 | 1 |

## 4. Results and Discussion

In this section, we discuss the performance measures used in this study and perform several evaluations. First, we compare four different versions of our proposed method (four different backbones including VGG16 [53], ResNet50 [54], DenseNet121 [55], and EfficientNetB0 [56]) against feature-based methods (HOG+SVM), off-the-shelf CNNs, and several recently-developed methods using two separate public datasets (NEH dataset released in this study and UCSD dataset). Second, we study the optimum number of feature maps to be merged together in order to achieve the best results. Third, we analyze the effect of gradual learning on improving the model's evaluation metrics. Fourth and last, Class Activation Maps (CAMs) are generated using the Grad-CAM method to visualize the key features used by the model for distinguishing AMD and normal cases.

### 4.1. Performance Measures

Classification performance of the evaluated models are obtained from the 3-class, and 4-class confusion matrix for the NEH and the UCSD dataset, respectively. The number of model parameters, runtime per epoch, loss function value, sensitivity, specificity, and accuracy are used for performance analysis on the first dataset (Table 6). For the second dataset, sensitivity, specificity, accuracy, and presence of a preprocessing step are compared (Table 7). Sensitivity and specificity measure the proportion of positives and negatives that are correctly classified, and accuracy is the percentage of correct predictions of the model.

In this section, we evaluated our proposed architecture performance using two separate datasets. For the first comparison, we implemented our proposed multi-scale CNN with different backbones (VGG, ResNet, DenseNet, and EfficientNet) and compared them against their corresponding base models, a feature-based model based on HOG feature extractor and SVM classifier, and several recently-proposed retinal OCT classification frameworks including one with a multi-scale CNN architecture [46]. Table 6 shows the average performance of all models in a five-fold cross-validation setup.

**Table 6.** Classification results of a 3-class classification problem on the NEH dataset published in this study. Performance measures are according to five-fold cross-validation.

| Model Description | Model Method | # Param (mil) | Runtime / epoch (sec) | Accuracy (%) | Sensitivity (%) | Specificity (%) | Weighted CCE Loss |
|---|---|---|---|---|---|---|---|
| Feature-based Method | HOG + SVM | - | - | 67.2 ± 3.7 | 66.9 ± 3.1 | 74.3 ± 2.5 | - |
| Base Deep Learning Models | VGG16[1] [58] | 28.3 | 110 | 91.6 ± 2.2 | 91.4 ± 2.0 | 95.6 ± 1.1 | 0.31 ± 0.11 |
| | ResNet50 [59] | 23.6 | 111 | 86.8 ± 2.0 | 86.4 ± 1.6 | 93.0 ± 0.9 | 0.40 ± 0.11 |
| | DenseNet121 [55] | 7.0 | 123 | 90.0 ± 1.4 | 89.7 ± 1.7 | 94.7 ± 0.8 | 0.31 ± 0.05 |
| | EfficientNetB0 [60] | 4.0 | 117 | 85.4 ± 2.6 | 84.5 ± 2.2 | 92.1 ± 1.3 | 0.40 ± 0.06 |
| Previous Studies | Kermany et al. [20] | 0.02 | 236[2] | 83.9 ± 1.7 | 82.9 ± 2.3 | 91.4 ± 1.0 | 0.42 ± 0.06 |
| | Kaymak et al. [42] | 58.3 | 109 | 80.2 ± 4.7 | 80.0 ± 4.4 | 89.4 ± 2.5 | 0.53 ± 0.11 |
| | Thomas et al. [46] | 2.5 | 112 | 68.5 ± 4.99 | 69.1 ± 4.3 | 83.8 ± 2.8 | 0.68 ± 0.07 |
| Proposed Structure with different Backbones | FPN-VGG16 | 21.6 | 167 | **92.0 ± 1.6** | **91.8 ± 1.7** | **95.8 ± 0.9** | **0.28 ± 0.11** |
| | FPN-ResNet50 | 31.1 | 176 | 90.1 ± 2.9 | 89.8 ± 2.8 | 94.8 ± 1.4 | 0.34 ± 0.08 |
| | FPN-DenseNet121 | 14.3 | 196 | 90.9 ± 1.4 | 90.5 ± 1.9 | 95.2 ± 0.7 | 0.31 ± 0.07 |
| | FPN-EfficientNetB0 | 12.7 | 181 | 87.8 ± 1.3 | 86.6 ± 1.8 | 93.3 ± 0.8 | 0.36 ± 0.05 |

As can be observed, using the FPN structure to create a multi-scale CNN results in a performance boost for all the tested backbones. The performance boost varies from 0.4% (from 91.6% ± 2.2% to 92.0% ± 1.6%, for the VGG16 model) to 3.3% (from 86.8% ± 2.0% to 90.1% ± 2.9%, for the ResNet50 model) in terms of accuracy. Also, all the proposed multi-scale CNN architectures achieve superior performance against feature-based (HOG+SVM), transfer learning-based method implemented by Kermany et al. [20], AlexNet model implemented by Kaymak et al. [42], and a multi-scale CNN structure proposed by Thomas et al. [46].

---

[1] Two dense layers of size 4096 are replaced with dense layers of size 512 in order to make the number of parameters comparable between base and FPN-based model.
[2] Input images are in 299 × 299 × 3 dimension, which is different than all other models having input dimensions of 224 × 224 × 3. This makes the runtime/epoch for this model uncomparable.

For the second comparison, the best-performing multi-scale CNN architecture from the last step (FPN-VGG16) was compared against off-the-shelf CNNs and multiple well-known retinal OCT classification frameworks that reported accuracies on the UCSD dataset [20]. Table 7 shows the overall performance for the 4-class classification problem using the UCSD dataset for four types of studies. In the first study, the FPN-VGG16 was compared against three off-the-shelf-CNNs using 1000 test images of the UCSD dataset (third/last version). In the second study, the FPN-VGG16 was compared against multiple previous studies that reported results on the same test of the UCSD dataset. In the third study, the FPN-VGG16 was compared against a study conducted by Das et al. [45] that was tested on the second version of the UCSD dataset. For this comparison, the FPN-VGG16 model was trained on the same dataset as in [45] and tested on the same images. In the fourth and last study, the FPN-VGG16 model was compared against two studies that conducted five-fold cross-validation on training images of the UCSD dataset. For this comparison, the FPN-VGG16 model was trained using five-fold cross-validation on the UCSD training dataset.

Table 7 shows the results for four types of studies discussed above. The results of the first study indicate the superior performance of the proposed FPN-VGG16 model against off-the-shelf CNN models, which emphasizes the effectiveness of having a multi-scale structure through feature combination. The second study demonstrates the superior performance of our model against several previous studies on retinal OCT classification [20], [42], [61]. However, the model proposed by Hassan et al. [51] shows slightly better results in terms of overall accuracy (up by 0.2% compared to our model). This is expected since the RAG-FW model [51] uses a preprocessing stage to crop the retina and has a hybrid CNN structure that benefits from additional information provided from a segmentation unit. For the third study, our model was compared against a multi-scale deep feature fusion (MDFF) model proposed by Das et al. [5]. This model takes advantage of a preprocessing block (consisting of retinal flattening, image cropping, and image normalization), multi-scale spatial pyramid decomposition (MSSP) to create multi-scale information of input images, and a classification block consisting of four CNNs. Our proposed multi-scale structure correctly classified 999 cases out of 1000 (the only incorrect classification was the detection of one drusen image as CNV), resulting in an accuracy of 99.9%, which is a

0.3% improvement over the study conducted by Das et al. [5]. This result shows the effectiveness of our proposed methodology, which enables end-to-end training with a single input image without any need to perform MSSP decomposition or to preprocess input data. In the last study, our multi-scale architecture was compared against two studies conducted by Fang et al. [16], [44]. One study proposes a feature fusion strategy to iteratively combine features in layers of CNNs [44], and the other one uses a lesion detection network (LDN) to generate an attention map and incorporates it into a classification framework [16]. Our proposed model achieved superior performance on the NEH dataset compared to both proposed methodologies.

Table 7. Classification results of a 4-class classification problem on the UCSD dataset.

| Study Num | Details | | Accuracy (%) | Sensitivity (%) | Specificity (%) | Preprocessing |
|---|---|---|---|---|---|---|
| Study #1 | Comparison against off-the-shelf CNNs on the UCSD dataset [20] (last version) | VGG16 [53] | 93.9 | 100 | 90.8 | ✗ |
| | | ResNet50 [54] | 96.7 | 99.6 | 94.8 | ✗ |
| | | EfficientNetB0 [56] | 95.0 | 99.8 | 91.4 | ✗ |
| | | FPN-VGG16 | 98.4 | 100 | 97.4 | ✗ |
| Study #2 | Comparison against four studies on the UCSD dataset (last version) | Kermany et al. [20] | 96.6 | 97.8 | 97.4 | ✗ |
| | | Kaymak et al. [42] | 97.1 | 98.4 | 99.6 | ✗ |
| | | Hwang et al. [7] | 96.9 | – | – | ✗ |
| | | Hassan et al. [51] | 98.6 | 98.27 | 99.6 | ✓ [1] |
| | | FPN-VGG16 | 98.4 | 100 | 97.4 | ✗ |
| Study #3 | Comparison against a study on the UCSD dataset (2nd version) | Das et al. [5] | 99.6 | 99.6 | 99.87 | ✓ |
| | | FPN-VGG16 | 99.9 | 100 | 99.8 | ✗ |
| Study #4 | Comparison against studies on the UCSD dataset using five-fold cross-validation (last version) | Fang et al. (JVCIR) [44] | 87.3 | 84.7 | 95.8 | ✗ |
| | | Fang et al. (TMI) [16] | 90.1 | 86.8 | 96.6 | ✓ [2] |
| | | FPN-VGG16 | 93.9 | 93.4 | 98.0 | ✗ |

---

[1] Input scan was first preprocessed through structure tensors to crop the retina and remove background information, and the image was then passed to a segmentation unit for lesion extraction.
[2] A lesion detection network (LDN) is first used to generate a soft attention map from the whole OCT image.

The performance of our proposed method could be further improved by incorporating preprocessing blocks (such as retinal cropping and flattening) and using more complex feature fusion styles [62]–[67]. However, the goal of this study was to demonstrate the power of feature fusion using a simple and understandable design (FPN) for classification. To our knowledge, this is the first study that investigates retinal OCT classification using FPN structures.

### 4.2. Choice of merged scales number

In this section, we aimed to find the optimum number of merging feature maps. The results are analyzed on the FPN-VGG16, the best-performing model on the NEH dataset.

VGG16 structure consists of five convolutional blocks, where each block has two or three convolutional layers. In Fig. 2, all five convolutional blocks are utilized and merged to build the final model. However, merging all blocks would not necessarily result in the best performance. To study the effect of feature fusion, we have run the models with five different fusion strategies. In the first setting, we only used the top convolutional block (scale $i = \{5\}$) for retinal pathology classification. In the second setting, we fused features of the last two convolutional blocks (scales $i = \{4,5\}$) and measured the performance. The other three settings include adding one more scale each time (scales $i = \{3,4,5\}, \{2,3,4,5\}, \{1,2,3,4,5\}$)

Table 8 presents the results for different combinations of merged scales. It can be observed from the results that fusing more feature maps increases the number of parameters for the model. Also, it can be seen that the best performance is achieved when using the top 3 scales of the FPN-VGG16 model ($i = \{3,4,5\}$). The results could be explained in two ways:

1. While later convolutional layers have strong semantics and low resolution, earlier layers have weak semantics and high resolution. Thus, it could be interpreted that fusing earlier convolutional blocks to the final structure would not benefit the whole model significantly.

2. The increase in the number of model parameters in higher scales (top-4 and top-5) increases the chance for overparameterization and overfitting.

Considering the reasons mentioned above, we can conclude that a trade-off should be found between the number of trainable parameters and fused feature maps. In this problem, the optimum point is found to be at the scale of $i = 3$.

**Table 8.** Average performance of models in a five-fold cross-validation setup for different combinations of merged scales.

| Model | | | Evaluation Metrics | | | |
|---|---|---|---|---|---|---|
| Encoder | Type | # Param (mil) | Accuracy (%) | Sensitivity (%) | Specificity (%) | Weighted CCE Loss |
| FPN-VGG16 | Top-1 | 16.2 | 91.1 ± 1.7 | 90.4 ± 2.1 | 95.2 ± 0.9 | 0.27 ± 0.07 |
| | Top-2 | 17.6 | 91.3 ± 1.2 | 90.9 ± 1.4 | 95.3 ± 0.8 | 0.30 ± 0.08 |
| | Top-3 | 19.0 | ==92.4 ± 2.4== | ==92.0 ± 2.5== | 95.9 ± 1.3 | ==0.24 ± 0.09== |
| | Top-4 | 20.3 | 92.3 ± 1.3 | 91.9 ± 1.4 | ==95.9 ± 0.8== | 0.25 ± 0.06 |
| | Top-5 | 21.6 | 92.0 ± 1.6 | 91.8 ± 1.7 | 95.8 ± 0.9 | 0.28 ± 0.11 |

### *4.3. Assessing the effect of gradual learning*

Deep learning models often require a large amount of training data to perform well as they have a huge number of parameters that need to be tuned by the learning algorithm. However, gathering large training sets in medical image analysis is a tedious and time-consuming process as it requires experts' skills, energy, and time. Transfer learning has proved to be an effective strategy in reducing the need for large-scale databases in order to obtain good performance with deep neural networks. The intuition behind transfer learning is that a model trained on a general and large-scale database could be used as a generic model of the visual world and a starting point for a model on the second task. Using a similar idea, we hypothesized that transferring the model's knowledge from a large-scale database in a related medical field could further enhance generalizability and be a good starting point for training the final model. Thus, we proposed a two-staged gradual learning strategy, where the model gradually adapts itself to classifying retinal OCT images. In the first stage, we loaded ImageNet weights to the FPN-VGG16's encoder part and used this pre-trained model as a starting point for the second stage. In the second stage, we fine-tuned the model on a large-scale public database consisting of more than 100k retinal OCT images [20]. This fine-tuned model was used as a starting point for classifying OCT images in the NEH database published in this study. It should be pointed out that the number of output neurons was matched to the classification problem (four neurons for training on the UCSD dataset and three neurons for training on the NEH dataset). The hypothesis is that this gradual adaptation using large-scale databases would help find a better local minimum for a non-convex training criterion.

To test our hypothesis, we have trained the FPN-VGG16 model using three procedures:

1. In the first procedure, we randomly initialized the weights and trained the model on the NEH dataset published in this study.

2. In the second procedure, we loaded ImageNet pre-trained weights on the encoder part of our FPN-VGG16 model and fine-tuned the model on the NEH dataset.

3. In the third procedure, we loaded ImageNet pre-trained weights in the first stage, fine-tuned the model on the UCSD dataset [20], and fine-tuned the model again on the NEH dataset.

The results are summarized in Table 9. Comparing strategies 1 and 2, we observed a 4.8% increase in overall accuracy (from 87.2% ± 2.5% to 92.0% ± 1.6%), which can be attributed to using Image-Net pre-trained weights as starting point to train the model on the NEH database. Comparing strategies 2 and 3, we observed another 1.4% increase in overall accuracy (from 92.0% ± 1.6% to 93.4% ± 1.4%), which can be attributed to the incremental effect of using knowledge in a related domain. The second stage of training seems to have provided a better starting point by guiding the learning algorithm towards better regions (i.e., basins of attractions). This is similar to the idea of curriculum learning [68], where the model starts with learning simpler concepts first (e.g., learning edges and shapes as in the ImageNet database), and then gradually expands its resources and learn more complex ones (e.g., learning lesion differences in retinal OCT images). The results demonstrated the effectiveness of the gradual learning strategy in finding better local minimum (lower CCE loss) of the non-convex training criterion and improving the generalizability of the model. Fig. 3 provides a training diagram for three strategies tested in this section.

**Table 9.** Average performance of models in a five-fold cross-validation setup for evaluating the effect of gradual learning on the Noor Eye Hospital dataset.

| Model | | Evaluation Metrics | | | |
|---|---|---|---|---|---|
| Encoder | Weights | Accuracy (%) | Sensitivity (%) | Specificity (%) | Weighted CCE Loss |
| FPN-VGG16 | Random Initialization | 87.2 ± 2.5 | 86.7 ± 2.6 | 93.1 ± 1.4 | 0.39 ± 0.14 |
| | ImageNet | 92.0 ± 1.6 | 91.8 ± 1.7 | 95.8 ± 0.9 | 0.28 ± 0.11 |
| | ImageNet + OCT | 93.4 ± 1.4 | 93.1 ± 1.7 | 96.5 ± 0.8 | 0.24 ± 0.08 |

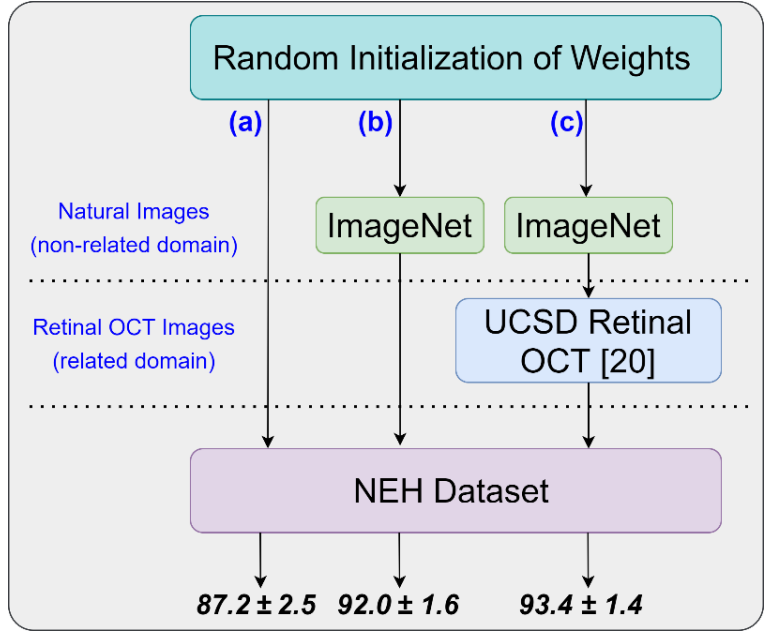

**Fig. 3.** Schematic presentation of three learning strategies. In strategy (a), the model was initialized with random weights and trained on the NEH dataset. This strategy achieved an accuracy of 87.2% ± 2.5%. In strategy (b), the model was pre-trained on the ImageNet database and fine-tuned on the NEH dataset. This strategy achieved an accuracy of 92.0% ± 1.6%. In strategy (c), the model was pre-trained on the ImageNet database, fine-tuned on the UCSD retinal OCT database, and fine-tuned again on the NEH dataset. This strategy achieved an accuracy of 93.4% ± 1.4%. The gradual learning strategy seems to have provided a better starting point by guiding the learning algorithm towards better regions.

### 4.4. Visualizing decision maps for the proposed multi-scale structure

Gaining insight into the model's key features to diagnose a pathology has significant importance to medical doctors and patients. Thus, in this study, we interpreted the results by plotting CAMs via the Grad-CAM technique [69] and discussed the effectiveness of our multi-scale CNN approach in improving the overall accuracy.

One major challenge with medical images is that regions of interest (ROIs) appear in varying scales. Thus, different-sized receptive fields would be needed in order to detect retinal pathologies. To achieve such an architecture, we proposed a multi-scale CNN structure based on the FPN design [52]. FPN's main applications are in object detection and semantic segmentation. However, we have modified their structure so that we would be able to take advantage of their multi-scale view in our classification problem.

In this section, we have plotted CAMs for the FPN-VGG16 model. One important upside of using this structure is its ability to detect pathologies in different sizes. Considering that the FPN-VGG16 model has five convolutional blocks, features were extracted at five different resolutions of 7 × 7, 14 × 14, 28 × 28, 56 × 56, and 112 × 112. Fig. 4 illustrates heatmaps of these five scales for a single CNV B-scan.

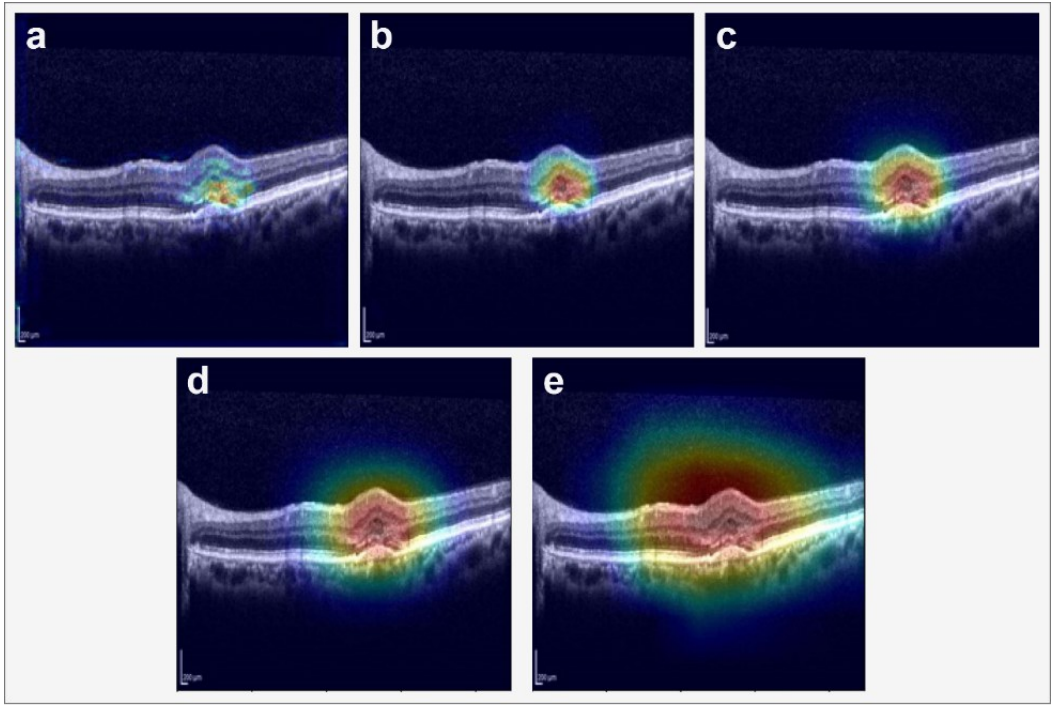

**Fig. 4.** Generated heatmaps using the Grad-CAM method for all fives scales of the FPN-VGG16 model. (a) scale 1 with a 112 × 112 output, (b) scale 2 with a 56 × 56 output, (c) scale 3 with a 28 × 28 output, (d) scale 4 with a 14 × 14 output, and (e) scale 5 with a 7 × 7 output.

There are two major benefits with the proposed multi-scale CNN structure:

1. Retinal pathologies that are not distinguishable on a single scale might be identified in higher/lower scales. Fig. 5 shows a drusen case where the last convolutional block was not able to localize the area of interest. The reason for this failure could be justified by the small size of drusen present in the OCT image, which made it difficult for the last convolutional block (with the coarsest resolution) to correctly locate the lesion area. However, the model correctly identifies the deposit in the retina associated with drusen when using a finer scale. This case was correctly classified as drusen with a high probability of 90.3%, showing the effectiveness of the proposed multi-scale structure.

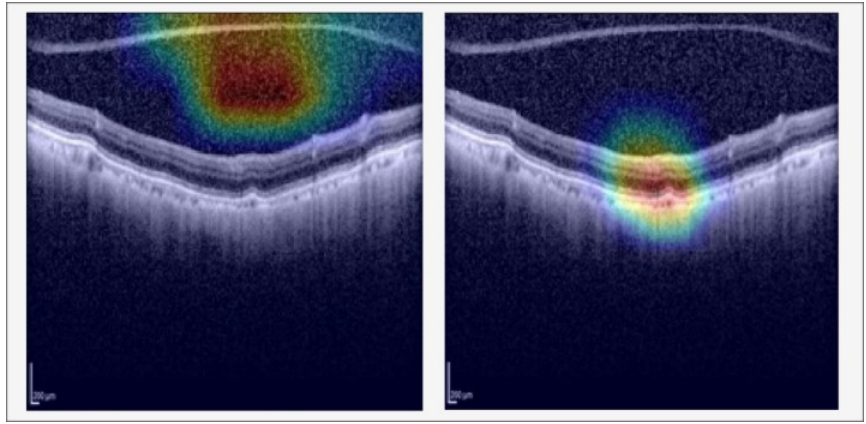

**Fig. 5.** Heatmaps for the first drusen case using fifth (left) and fourth (right) convolutional block output for the FPN-VGG16 model. As can be observed, the last convolutional block was not able to localize the area of interest. However, the model correctly identifies the deposit in the retina associated with drusen when using the output from a finer scale. Correct classification of this case as drusen with a probability of 90.3% shows the effectiveness of the proposed multi-scale structure.

2. The multi-scale approach provides the expert with a finer look into the model's decision-making process. While the fifth block of the VGG16 model in the FPN structure provides a $7 \times 7$ heatmap using the Grad-CAM method, the fourth block provides a $14 \times 14$ heatmap, having four times more resolution than the fifth block (twice more resolution in the x and y-direction). Fig. 6 illustrates a drusen case where the last convolutional layer could not precisely pinpoint macular lesion. On the other hand, the finer-scale block provides the expert with a more delicate look into the key features used by the model in classifying this case as drusen.

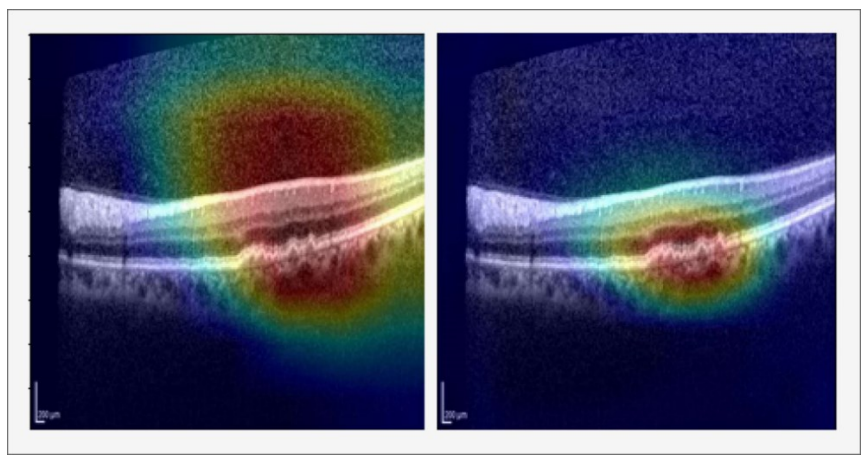

**Fig. 6.** Heatmaps for the second drusen case using fifth (left) and fourth (right) convolutional block output for the FPN-VGG16 model. The last convolutional block's heatmap is unable to precisely locate the macular lesion. However, the fourth convolutional block provides experts with a finer look into the model's decision-making process for classifying this B-scan as drusen.

## 5. Conclusion

In this paper, we proposed a multi-scale automated method for classifying AMD-related retinal pathologies. The two main contributions of this study were: (a) designing a multi-scale CNN architecture through feature fusion based on FPN architecture, enabling end-to-end training and reducing computational complexity compared to the parallel use of multiple CNNs, and (b) additional performance enhancement using a two-staged (gradual) learning strategy. The advantage of the proposed feature fusion strategy is making use of a single CNN, leveraging the pyramidal shape of the feature hierarchy to create a multi-scale view. The results demonstrated the superior performance of the proposed structure compared to several well-known retinal OCT classification frameworks. The improvements observed from all FPN-based structures when compared to their base models prove the effectiveness of the feature fusion strategy used in this study. Besides, we observed that tuning the number of fusing feature maps in a multi-scale structure would help in improving the model's performance. Also, gradual learning has proved to be an effective method in finding better local minimum (lower CCE loss) of the non-convex training criterion and improving the generalizability of the model. In the end, qualitative evaluation of generated heatmaps via the Grad-CAM technique proved the added value of a multi-scale structure, making this model a convenient screening tool for reducing the burden on healthcare centers and assisting ophthalmologists in making better diagnostic decisions.

**Conflicts of Interest:** None declared.

**Funding/Support:** None declared.

**Informed Consent:** Informed consent was waived because of the retrospective nature of this study. There were potentially no marks/features and no patient identifiers in the images.